\documentclass{article}

\usepackage{arxiv}

\usepackage[utf8]{inputenc} 
\usepackage[T1]{fontenc}    
\usepackage{hyperref}       
\usepackage{url}            
\usepackage{booktabs}       
\usepackage{amsfonts}       
\usepackage{nicefrac}       
\usepackage{microtype}      
\usepackage{lipsum}         
\usepackage{graphicx}
\usepackage{natbib}
\usepackage{doi}

\usepackage{comment}
\usepackage{booktabs}
\usepackage{multirow}
\usepackage{graphicx}
\usepackage{subcaption}
\usepackage{colortbl}  
\usepackage{mwe} 
\usepackage{amsmath}
\usepackage{amssymb}
\usepackage{float}

\usepackage{cleveref}
\usepackage{xcolor}

\title{Test Time Optimized Generalized AI-based Medical Image Registration Method}

\date{}

\newif\ifuniqueAffiliation

\ifuniqueAffiliation 
\author{ \href{https://orcid.org/0000-0000-0000-0000}{\includegraphics[scale=0.06]{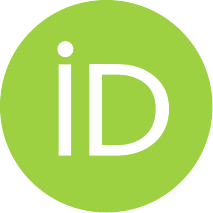}\hspace{1mm}David S.~Hippocampus}\thanks{Use footnote for providing further
		information about author (webpage, alternative
		address)---\emph{not} for acknowledging funding agencies.} \\
	Department of Computer Science\\
	Cranberry-Lemon University\\
	Pittsburgh, PA 15213 \\
	\texttt{hippo@cs.cranberry-lemon.edu} \\
	\And
	\href{https://orcid.org/0000-0000-0000-0000}{\includegraphics[scale=0.06]{orcid.pdf}\hspace{1mm}Elias D.~Striatum} \\
	Department of Electrical Engineering\\
	Mount-Sheikh University\\
	Santa Narimana, Levand \\
	\texttt{stariate@ee.mount-sheikh.edu} \\
}
\else
\usepackage{authblk}

\author[1]{%
	Sneha Sree C\thanks{\texttt{snehasree.c@gehealthcare.com}}%
}
\author[1]{%
	Dattesh Shanbhag\thanks{\texttt{dattesh.shanbhag@gehealthcare.com}}%
}
\author[1]{%
	Sudhanya Chatterjee\thanks{\texttt{sudhanya.chatterjee@gehealthcare.com}}%
}
\affil[1]{GE HealthCare, Bangalore, India}
\fi

\renewcommand{\shorttitle}{Test time optimized DL based medical image registration}

\hypersetup{
pdftitle={Test Time Optimized Generalized AI-based Medical Image Registration Method},
pdfsubject={q-bio.NC, q-bio.QM},
pdfauthor={Sneha Sree C., Dattesh Shanbhag, Sudhanya Chatterjee},
pdfkeywords={Non-Rigid Registration (NRR), Test Time Optimization (TTO), Deep Learning.},
}

\begin{document}
\maketitle

\begin{abstract}
Medical image registration is critical for aligning anatomical structures across imaging modalities such as computed tomography (CT), magnetic resonance imaging (MRI), and ultrasound. Among existing techniques, non-rigid registration (NRR) is particularly challenging due to the need to capture complex anatomical deformations caused by physiological processes like respiration or contrast-induced signal variations. Traditional NRR methods, while theoretically robust, often require extensive parameter tuning and incur high computational costs, limiting their use in real-time clinical workflows. Recent deep learning (DL)-based approaches have shown promise; however, their dependence on task-specific retraining restricts scalability and adaptability in practice. These limitations underscore the need for efficient, generalizable registration frameworks capable of handling heterogeneous imaging contexts. In this work, we introduce a novel AI-driven framework for 3D non-rigid registration that generalizes across multiple imaging modalities and anatomical regions. Unlike conventional methods that rely on application-specific models, our approach eliminates anatomy- or modality-specific customization, enabling streamlined integration into diverse clinical environments.
\end{abstract}


\section{Introduction}
\label{sec:introduction}
Medical image registration is a fundamental technique used to align two or more images of the same anatomical region, acquired at different time points or through different imaging modalities~\cite{hill2001medical, maintz1998survey}. Its primary goal is to establish spatial correspondence between images, enabling the integration of information from multiple sources for improved analysis. This process plays a critical role in various clinical applications, including treatment planning, disease monitoring, surgical guidance, and multi-modal image fusion~\cite{sotiras2013deformable}.

Image registration spatially aligns a moving image to a reference or fixed image. It uses (i) a transformation model (e.g., warp field), (ii) a similarity metric to assess alignment, and (iii) an optimization method that updates parameters to maximize similarity~\cite{oliveira2014medical}.
Transformation models can be rigid, affine, or deformable to capture complex anatomical variations, commonly referred to as non-rigid registration (NRR). While rigid and affine registration can be efficiently performed using analytical approaches~\cite{avants2009ants}, NRR is computationally challenging due to the complexity of deformation models and the iterative nature of optimization. Deep learning (DL)-based registration methods address these limitations by leveraging learned features to predict transformations directly, significantly reducing computation time for NRR~\cite{balakrishnan2019tmi, REITHMEIR2026103854}.
Recent advances in deep learning (DL)-based medical image registration have shown significant promise. Unsupervised frameworks such as \textit{VoxelMorph}~\cite{balakrishnan2019tmi} and \textit{DeepReg}~\cite{Fu2020} learn spatial transformations directly from data, enabling faster and more flexible registration. 
\textit{GroupRegNet}~\cite{zhang2021groupregnet} performs groupwise one-shot 4D registration, removing fixed references and reducing inference time. \textit{IIRP-Net}~\cite{ma2024iirpnet} uses residual pyramids and adaptive stopping without extra training. \textit{multiGradICON}~\cite{demir2024multigradicon} and \textit{uniGradICON}~\cite{tian2024unigradicon} offer zero-shot, cross-anatomy, cross-modality registration.
Despite these innovations, generalization to unseen domains remains a major challenge, as most models rely heavily on training data distributions and struggle with domain shifts encountered in clinical practice~\cite{REITHMEIR2026103854}.
We present a 3D DL registration framework that generalizes across applications. We train on multi‑modal‑like synthetic data~\cite{Hoffmann_2022}. For each target, we use test‑time optimization (TTO) with deformation‑field regularization. To offset TTO latency, we apply knowledge distillation~\cite{hinton2015distilling} to reduce model size without loss.
We evaluate on 4D DCE‑MRI motion correction, MRI–CT alignment, pre/post‑contrast CT with large deformations, and inter‑subject brain MRI. Visual and quantitative metrics show high accuracy. All experiments use the same base model and TTO.

\section{Method}
\label{sec:method}
\begin{figure}[htb]
	\centering
	\includegraphics[width=1.0\textwidth]{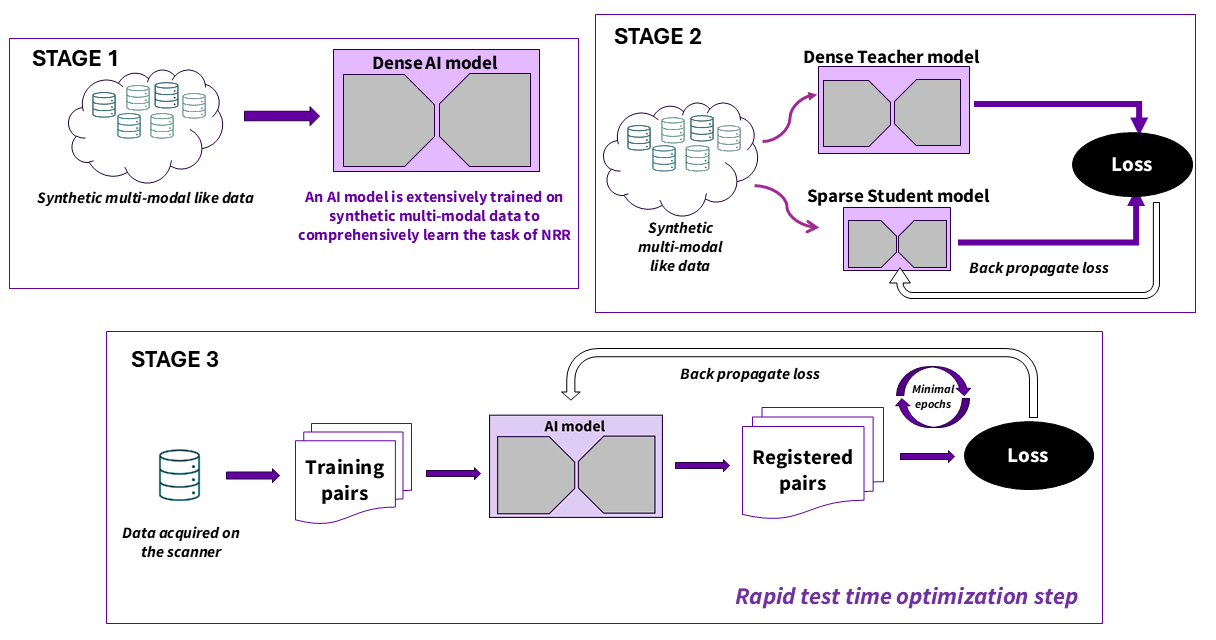}
	\caption{Overview of the three stage pipeline of the proposed method.}
	\label{fig:two_stage_approach}
\end{figure}

We propose a three-stage framework for DL based medical image registration. The overall pipeline is shown in Figure~\ref{fig:two_stage_approach}. The three stages shown here are discussed in Section~\ref{subsec:Stage1ModelPreTraining}, \ref{subsec:Stage2KD}, \ref{subsec:Stage3TTO}.

\subsection{Stage-1: Model Pretraining}\label{subsec:Stage1ModelPreTraining}
\label{sec:Generalized Model Pretraining}
Let $I_f$ and $I_m$ denote the fixed and moving 3D images, respectively. The objective of image registration is to estimate a spatial transformation that aligns $I_m$ to $I_f$. We adopt a learning-based approach where a 3D convolutional neural network (CNN), parameterized by $\theta$, predicts a dense displacement field (DDF): $\mathbf{u} = \mathcal{F}_\theta(I_f, I_m)$, where $\mathbf{u} : \Omega \subset \mathbb{R}^3 \rightarrow \mathbb{R}^3$ represents voxel-wise displacements. The transformation $\phi$ applied to a spatial location $\mathbf{x}$ is defined as: $\phi(\mathbf{x}) = \mathbf{x} + \mathbf{u}(\mathbf{x}).$
This formulation enables non-rigid registration by allowing local deformations across the entire image domain.

\subsubsection{Network Architecture}
\label{sec:network architecture dense}
\begin{figure}[htb]
	\centering
	\includegraphics[width=1.0\textwidth]{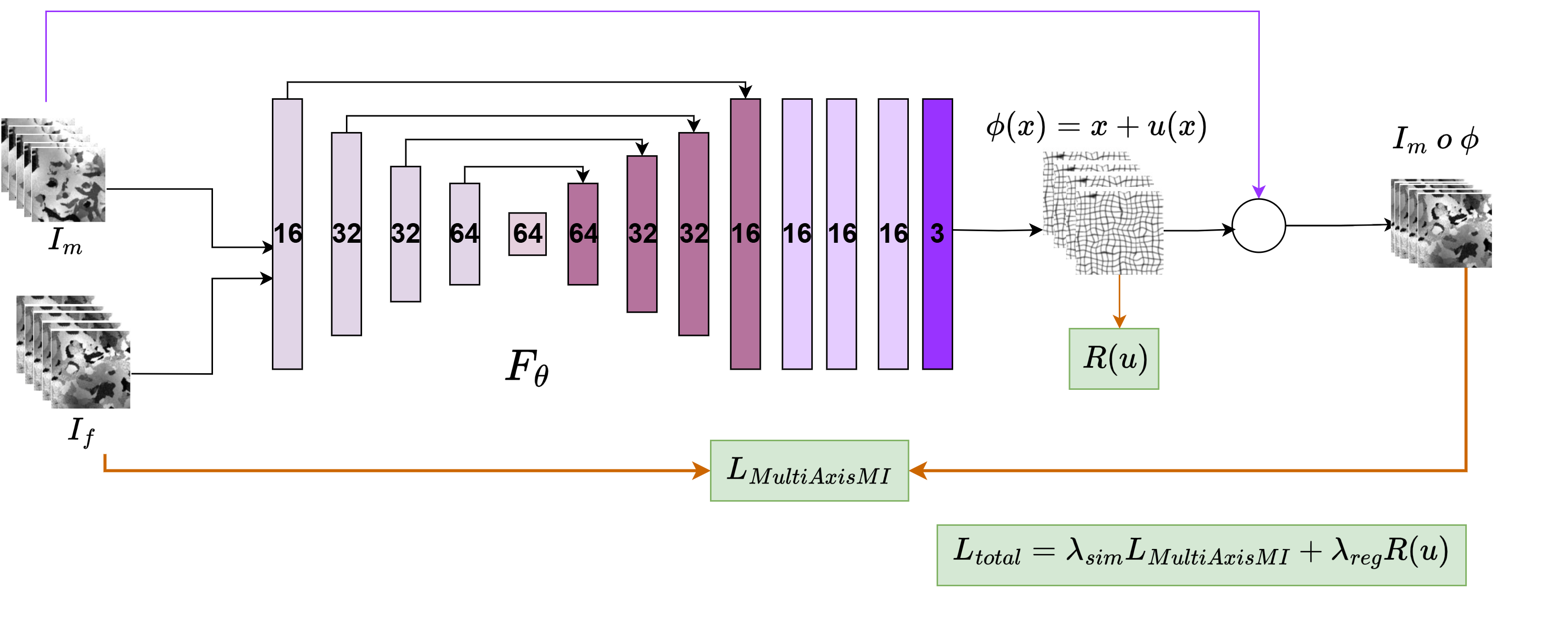}
	\caption{3D U-Net based registration architecture.}
	\label{fig:Teacher Network}
\end{figure}
Our registration framework is built upon a 3D U-Net architecture ~\cite{cciccek20163d}, optmized for learning dense deformation fields.\\ 
\textbf{Encoder-Decoder Structure}: The DL network is illustrated in Figure~\ref{fig:Teacher Network}. The encoder consists of four hierarchical levels with feature channels of 16, 32, 32, and 64. Each level applies a $3 \times 3 \times 3$ convolution (stride 1, padding 1) followed by LeakyReLU activation ($\alpha = 0.2$), with downsampling performed via $2 \times 2 \times 2$ max-pooling.\\ 
\textbf{Displacement Field Prediction}: Following the decoder, the network employs a sequence of full-resolution refinement layers. Three convolutional blocks, each containing a $3 \times 3 \times 3$ convolution with 16 output channels followed by LeakyReLU activation ($\alpha = 0.2$), progressively refine the feature representation while maintaining spatial resolution, yielding  the displacement field as: $\mathbf{u} \in \mathbb{R}^{H \times W \times D \times 3}$.\\
\textbf{Spatial Transformation}: The moving image $I_m$ is warped to align with the fixed image $I_f$ using the predicted deformation field $\phi$ through trilinear interpolation:$I_m \circ \phi(\mathbf{x}) = I_m(\phi(\mathbf{x})) \quad \forall \mathbf{x} \in \Omega$ where $\Omega$ denotes the image domain.\\ 
\textbf{Loss Function}: The network is trained on synthetically generated image pairs using a composite loss that combines image similarity with deformation regularization:
\begin{equation}
	\mathcal{L}_{\text{total}} = \lambda_{\text{sim}} \cdot \mathcal{L}_{\text{MultiAxisMI}}(I_f, I_m \circ \phi) + \lambda_{\text{smooth}} \cdot \mathcal{R}(\mathbf{u}),
	\label{eq:total_loss}
\end{equation}
where $\lambda_{\text{sim}}$ and $\lambda_{\text{smooth}}$ are weighting coefficients, $I_m \circ \phi$ denotes the warped moving image.
The similarity term employs the Multi-Axis Mutual Information (MultiAxisMI) loss (see Appendix~\ref{sec:appendix-MultiAxisMI}, Eq.~\ref{eq:multi_axis_mi}), which captures alignment across depth, height, and width. To ensure anatomically plausible transformations, the regularization term penalizes abrupt spatial variations and hence enforcing smoothness constraint on the deformation field, $\mathcal{R}(\mathbf{u}) = \|\nabla \mathbf{u}\|^2,$ where $\nabla \mathbf{u}$ denotes spatial gradients of the displacement field.
\subsection{Stage-2: Knowledge Distillation (KD) for Efficient Deployment}\label{subsec:Stage2KD}
\label{sec:kd section}
KD is a widely used technique for compressing deep learning models by transferring knowledge from a large, high-capacity teacher network to a smaller student network while maintaining accuracy of the target task~\cite{hinton2015distilling}. To enable rapid test time optimization, this is a crucial step.\\
\textbf{Student Network Architecture}: The student network adopts a U-Net architecture similar to the teacher (refer Appendix~\ref{sec:appendix-KD-framework}, Figure ~\ref{fig:KD framework}). It employs a reduced channel configuration for efficiency: encoder with 16, 24, 24, and 32 channels and a 32-channel bottleneck; decoder mirrors this with 32, 24, 24, and 16 channels. This compression lowers the parameter count from ~0.6M to ~0.23M, achieving nearly a three-fold reduction in model size.\\
%
\textbf{Distillation Objective}: For KD we employ output-based distillation. The student network learns to approximate the displacement fields predicted by the teacher, ensuring that the deformation patterns and spatial correspondences captured by the teacher are effectively transferred.
Let $\mathbf{u}_T$ and $\mathbf{u}_S$ denote the displacement fields predicted by the teacher and student networks, respectively: $\mathbf{u}_T = \mathcal{F}_T(I_f, I_m), \quad \mathbf{u}_S = \mathcal{F}_S(I_f, I_m)$, with corresponding transformations: $
\phi_T(\mathbf{x}) = \mathbf{x} + \mathbf{u}_T(\mathbf{x})$, $ \phi_S(\mathbf{x}) = \mathbf{x} + \mathbf{u}_S(\mathbf{x})$. The total student training objective is formulated as:
\begin{equation*}
	\mathcal{L}_{\text{KD}} = \lambda_{\text{dist}} \cdot \mathcal{L}_{\text{MultiAxisMI}}(I_m \circ \phi_T, I_m \circ \phi_S) + \lambda_{\text{sim}} \cdot \mathcal{L}_{\text{MultiAxisMI}}(I_f, I_m \circ \phi_S) + \lambda_{\text{smooth}} \cdot \|\nabla \mathbf{u}_S\|^2
\end{equation*}
where, $\mathcal{L}_{\text{MultiAxisMI}}$ computes slice-wise mutual information across three directions (refer Appendix~\ref{sec:appendix-MultiAxisMI} equation~\ref{eq:multi_axis_mi}), $\|\nabla \mathbf{u}_S\|^2$ enforces smoothness in the predicted displacement field.

\subsection{Stage-3: Test-Time Optimization (TTO) at inference}\label{subsec:Stage3TTO}
During inference, the model undergoes a self-supervised fine-tuning phase to adapt to the domain-specific characteristics of the test data. This process, referred to as \textit{Test-Time Optimization (TTO)}. It involves optimizing the student network parameters $\theta$ for a small number of epochs, the optimization objective is expressed as:
\begin{equation}
	\theta^{*} = \arg\min_{\theta} \sum_{(I_f,I_m)} \mathcal{L}_{\text{TTO}}(I_f,I_m;\theta),
	\label{eq:TTO}
\end{equation}
where $\mathcal{L}_{\text{TTO}}$ is the loss at TTO and has the following terms.
\begin{itemize}
	\item \textbf{Image Similarity Term:} The similarity between the fixed image $I_f$ and the warped moving image $I_m \circ \phi$ is enforced using a combination of 3D Multi-Axis MI (Appendix~\ref{sec:appendix-MultiAxisMI} Equation~\ref{eq:multi_axis_mi}) and 3D Normalized Cross-Correlation (NCC) (Appendix~\ref{sec:appendix-ncc} Equation~\ref{eq:ncc_loss}): $\mathcal{L}_{\text{sim}} = \mathcal{L}_{\text{MultiAxisMI}}(I_f, I_m \circ \phi) + \mathcal{L}_{\text{NCC}}(I_f, I_m \circ \phi)$.
	
	\item \textbf{Smoothness Term:} The gradient-based regularization $\|\nabla \mathbf{u}\|^2$ is used to ensure smooth deformation field.
	\item \textbf{Divergence Regularization:} To further improve anatomical plausibility and prevent non-physical deformations, we introduce penalties on the divergence of the displacement field $\mathbf{u}$: $\mathcal{R}_{\text{div}}(\mathbf{u}) = \alpha_{\text{div}} \|\nabla \cdot \mathbf{u}\|^2$, where $\nabla \cdot \mathbf{u}$ measures local volume changes (encouraging near-incompressibility). The coefficient $\alpha_{\text{div}}$ controls the strength of this constraint.  This loss prevents tearing-like unrealistic effects in the registered image.
\end{itemize}
Hence, the TTO loss is given by: $\mathcal{L}_{\text{TTO}} = \lambda_{\text{sim}} \cdot \mathcal{L}_{\text{sim}} + \lambda_{\text{smooth}} \cdot \|\nabla \mathbf{u}\|^2 + \alpha_{\text{div}} \|\nabla \cdot \mathbf{u}\|^2$.

\subsubsection{Implementation Details}
\textbf{Framework and Hardware:} All models were implemented in PyTorch (v2.4.1+) and trained on NVIDIA RTX 6000 GPUs (48 GB VRAM).\\
\textbf{Teacher Network:} The teacher model was trained for 300 epochs on synthetic image pairs generated on-the-fly (500 pairs per epoch). Adam optimization. Learning rate of $1e{-4}$.\\
\textbf{Student Network:} After teacher convergence, the student model was trained using the knowledge distillation framework (Section~\ref{sec:kd section}) on the  synthetic dataset for 500 epochs.\\
\textbf{TTO at inference:} The pre-trained model was fine-tuned on target data. Stopping Criteria for TTO: 100 epochs or one minute of training, whichever is reached earlier. During the TTO step, data is resampled as needed to the nearest resolution of $2^4$ for $H \times W \times D$, ensuring compatibility with the U-Net-based model architecture.

\subsection{Experiments}
\label{subsec:experiments}
We evaluate the proposed method across multiple clinically relevant scenarios involving different imaging modalities, imaging contrasts (dynamic imaging included). In addition to visual confirmation of alignment, efforts are made to assess the performance using quantitative metrics. 
We introduce Patch‑wise MI Map (PMM) for registration evaluation. At each voxel, we extract a 3D patch and compute normalized mutual information (NMI) between the corresponding reference and target patches. Computational details are provided in Appendix~\ref{sec:appendix-patchwiseMI}. Note that, in multi-modal and multi-contrast applications, the values will not reach a value of 1. Across most experiments, we compare the proposed method against well established ANTs registration~\cite{avants2009ants}.

\subsubsection{Dynamic Contrast-Enhanced MRI (DCE-MRI)}\label{subsubsec:expdcemri}
DCE‑MRI acquires multiple 3D volumes over time (4D, 3D+t) to capture exogenous contrast dynamics. Respiratory motion--dominated by diaphragm motion in liver scans--misaligns frames and hinders reliable pharmaco-kinetic uptake quantification. We correct motion by registering all temporal phases to a reference phase using the proposed method. This restores alignment and enables accurate contrast‑uptake curves. Because image contrast changes markedly across time, the task is effectively a multi‑contrast registration problem.\\
\textbf{Dataset and Experiment description}: The MRI data was obtained on a 1.5T GE SIGNA EXCITE system, 3D EFGRE, TE/TR=1.12/4.8ms, matrix=\(256 \times 256 \times 32\), 30 bolus phase volumes (temporal frames), FA=15, FOV = 450$mm^2$. The middle time point is selected as the fixed reference volume, and the remaining 29 frames are treated as moving volumes. TTO is performed using the dense teacher model and the KD student model.\\
\textbf{Analysis}: To understand the alignment of structures over the temporal frames, the contrast uptake curves are plotted in a region of interest (ROI) at liver dome (region highly susceptible to motion). Additionally, we trained a comparison model, \textit{Application Specific AI model}, using the same dense teacher network and loss function as the TTO-based approach exclusively for liver DCE-MRI (500 epochs, 261 paired volumes). Achieving comparable performance between this model and the TTO-based method would suggest that TTO maintains structural alignment accuracy while providing enhanced generalizability.


\subsubsection{MRI–CT Multi-Modal Registration}\label{subsubsec:expmrictreg}
MRI and CT are different imaging modalities based on different principles. Consequently, tissue contrast varies significantly between modalities, for example, in MRI signal in bone is negligible but exhibits high intensity in CT. Due to these non-trivial intensity relationships, MRI–CT alignment is truly a multi-modal registration problem.\\
\textbf{Dataset and Experiment description}: We evaluate the proposed method on paired pelvic MRI and CT volumes acquired from the same subject. Each volume originally has dimensions of \(256 \times 256 \times 120\) voxels. The MRI and CT image are treated as the fixed and moving image respectively. Prior to this, the images are affine registered~\cite{avants2009ants}. The volumes are resampled to \(256 \times 256 \times 128\) and NRR is then performed using the proposed TTO approach. After registration, results are resampled back to the native resolution for evaluation.\\
\textbf{Analysis}: A visual comparison and PMM analysis were performed to demonstrate the alignment post registration using proposed method.

\subsubsection{Pre- and Post-Contrast CT Registration}\label{subsubsec:prepostct}
Pre-contrast and post-contrast CT scans are routinely acquired for diagnostic evaluation across anatomical regions. For a comparative analysis, it is important for the images to be aligned to each other.\\
\textbf{Dataset and Experiment Description}: TCIA datasets are used for this experiment. \textit{Data-1 (Thorax)} \cite{Bakr2017NSCLC} (Case ID: AMC-015), \textit{Data-2 (Abdomen)} \cite{Erickson2016TCGAUCEC} (Case ID: TCGA-DI-A2QT), \textit{Data-3 (Cardiac)} \cite{APOLLO2024VAREPOP} (Case ID: AP-26JK), and \textit{Data-4 (Liver)} \cite{Moawad2021HCC} (Case ID: HCC\_023). For all datasets pre- and post-contrast scans were considered as fixed and moving respectively. Before applying proposed DL based NRR, an affine alignment was performed. All the volumes were resampled to size of \(256 \times 256 \times 128\). After inference, the registered outputs were resampled back to their original resolution for evaluation.\\
\textbf{Analysis}: TotalSegmentator~\cite{wasserthal2023totalsegmentator} was used to segment the primary organ for pre-contrast image, post-contrast image before registration and post-contrast image after registration. Alignment between the segmented organs were quantified using Dice and Intersection-over-Union (IoU) scores.

\subsubsection{Cross-Contrast Brain MRI Registration}\label{subsubsec:expneuroreg}
Brain MRI studies often include multiple contrast sequences (e.g., T1-weighted, T2-weighted, Fluid-Attenuated Inversion Recovery (FLAIR)), and accurate registration across these contrasts can be important in comparative tasks.\\
\textbf{Dataset and Experiment Description}: In this experiment, we consider three experiments: Inter-subject and same contrast ($n=24$ cases), Inter-subject and cross-contrast ($n=24$ cases), and Same-subject but cross-contrasts ($n=8$ cases). These experiments cover all variations: registration across shape and contrast variations. The two contrasts considered here are T2-w (FLAIR) and T2-w Fast Spin Echo (FSE) MRI data. The fluid suppression in FLAIR images provide sufficient contrast variations between the images. The T2 FLAIR volume has dimensions \(512 \times 512 \times 22\) voxels with spacing \(0.4297 \times 0.4297 \times 6\) mm, and the T2-w FSE volume has similar dimensions and spacing. As a pre-processing step, both volumes are resampled to 1mm isotropic spacing. After registration, the outputs are resampled back to their native spacing and dimensions for evaluation.\\
\textbf{Analysis}: In addition to visual confirmations, PMM were computed and mean PMM values were obtained inside brain mask to quantify registration performance over entire population.

\section{Results and Discussion}
We present the outcomes of the proposed TTO framework across the experimental scenarios described in Section~\ref{subsec:experiments}. The registered output obtained using the dense teacher model is referred to as \textbf{DL Reg}, the output from the knowledge-distilled student model is denoted as \textbf{DL KD Reg}, registration output using ANTs is referred to as \textbf{ANTs Reg}.
\subsection{Dynamic Contrast Enhanced MRI (DCE-MRI)}
\begin{figure}[htb]
	\centering
	\caption{Left to right for Figure~\ref{fig:DCEMRIROIimages} show moving, registered using application specific AI model, DL Reg, and DL KD Reg. The contrast uptake curves in Figure~\ref{fig:ContrastUptakeCurvesDCEMRI} indicate expected correction post registration. Higher PMM values observed post-registration (Figure~\ref{fig:PMMDCEMRI}) indicate improved structural alignment.}
	\label{fig:example2}
	
	\begin{subfigure}{0.65\textwidth}
		\centering
		\includegraphics[width=\linewidth]{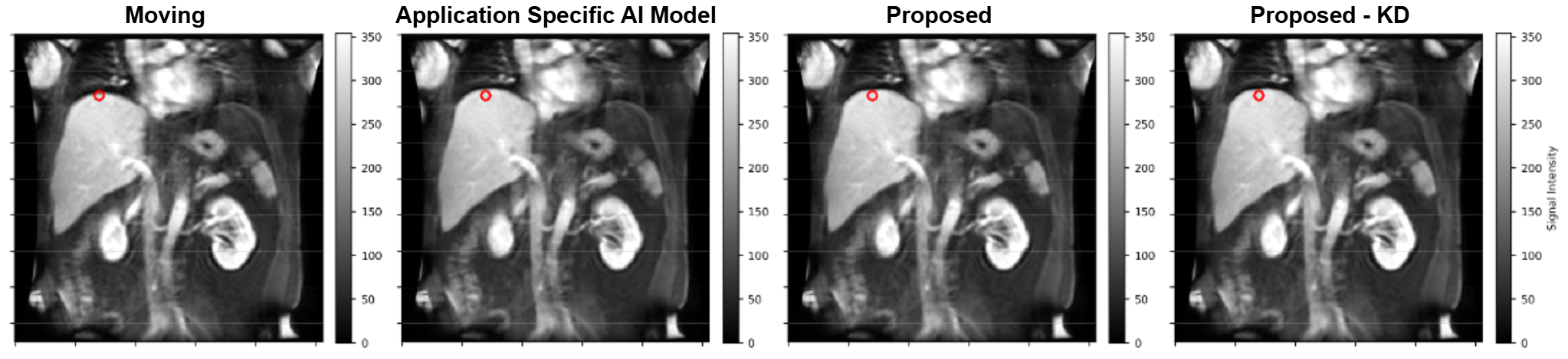}
		\caption{}
		\label{fig:DCEMRIROIimages}
	\end{subfigure}
	\hfill
	\begin{subfigure}{0.30\textwidth}
		\centering
		\includegraphics[width=\linewidth]{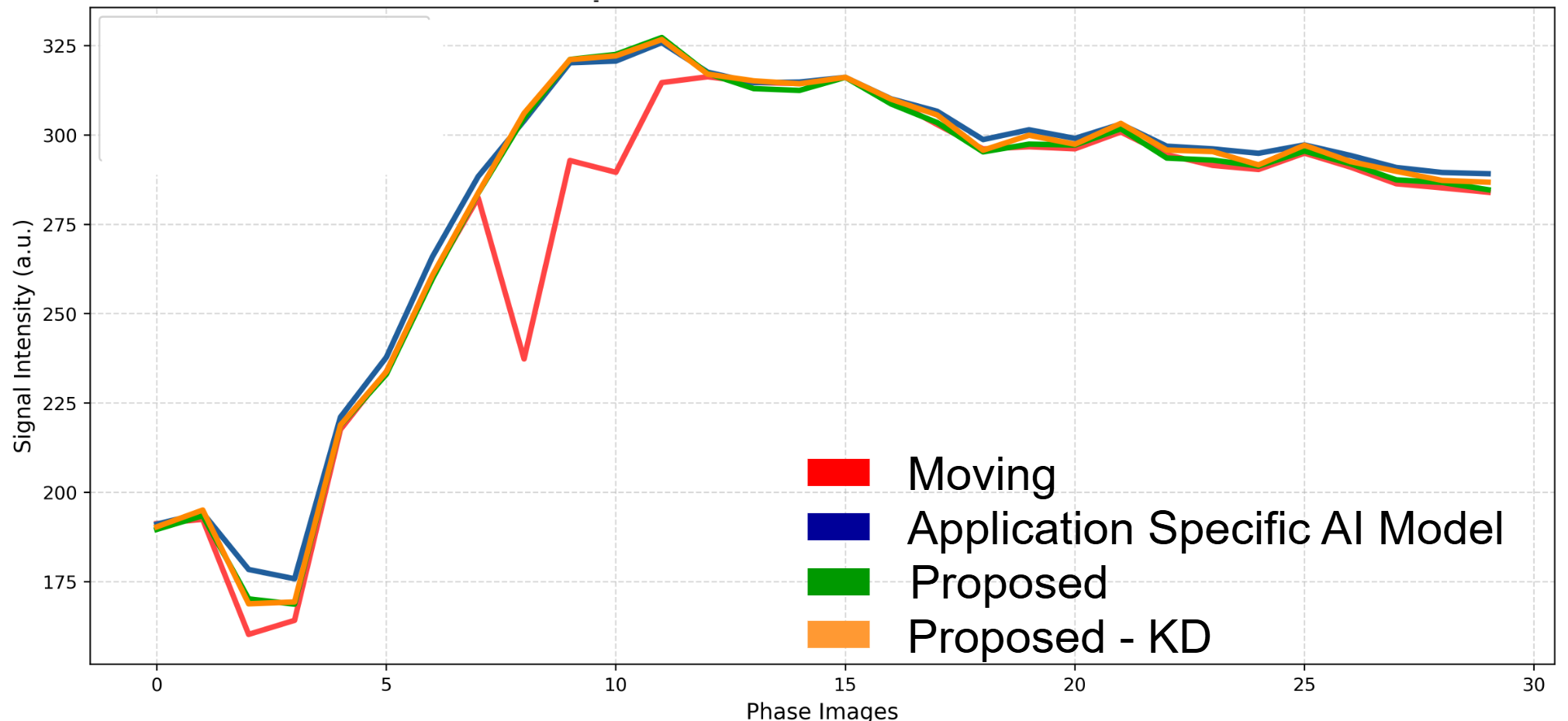}
		\caption{}
		\label{fig:ContrastUptakeCurvesDCEMRI}
	\end{subfigure}
	
	\medskip
	
	\begin{subfigure}{0.55\textwidth}
		\centering
		\includegraphics[width=\linewidth]{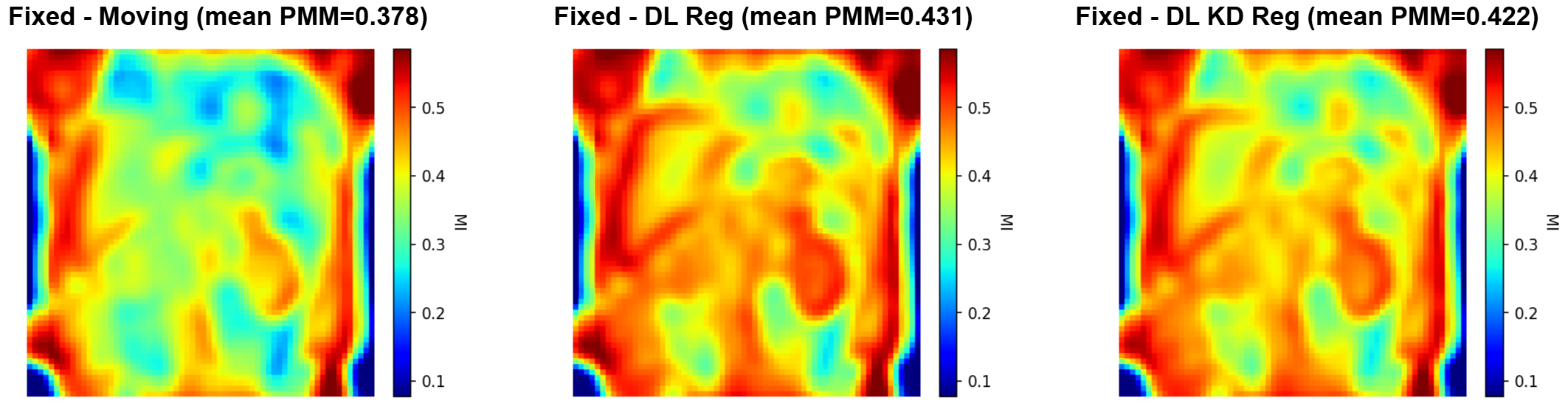}
		\caption{}
		\label{fig:PMMDCEMRI}
	\end{subfigure}
\end{figure}

The contrast uptake curves for red ROI in Figure~\ref{fig:DCEMRIROIimages} are shown in Figure~\ref{fig:ContrastUptakeCurvesDCEMRI}.
The sudden drop and spike in the contrast uptake curve (red) for moving image render them physiologically implausible. Post registration using the proposed method, both DL Reg (green curve) and DL KD Reg (orange curve), have corrected such abberations in the contrast uptake curve. The contrast uptake curves using proposed method matches the contrast uptake curve post correction using the application specific model (blue curve). This shows that the proposed generalizeable method performs as good as an AI model specifically trained for the application (hence, no compromise in performance). The green and orange curve matching each other indicates that the miniaturized KD model performance is equivalent to the dense model.
Image alignment post registration was evaluated using PMMs (see Appendix~\ref{sec:appendix-patchwiseMI}). Patch-wise normalized MI was computed for the image pairs: (\textit{fixed, moving}), (\textit{fixed, DL Reg}), and (\textit{fixed, DL KD Reg}) on 3D patches of size $16 \times 16 \times 16$ voxels with a stride of 4. The PMMs were subsequently averaged across timepoints for each slice to obtain a mean PMM. The mean PMMs for the target slice is shown in Figure~\ref{fig:PMMDCEMRI}.
The aggregated mean PMM values were: \textit{fixed–moving} = 0.3801, \textit{fixed–DL Reg} = 0.4291, and \textit{fixed–DL KD Reg} = 0.4195. The increased values indicate higher alignment post registration. This is evident from increased values in air pockets around the liver, kidney and spleen regions.
Visual assessments were also performed by overlaying the image pairs pre- and post registration using the proposed method (refer Appendix~\ref{sec:appendixQualityDCE} Figure~\ref{fig:overlayDCEMRI}).

%
\subsection{MRI-CT Multi-Modal Registration}
\begin{figure}[htb]
	\centering
	\caption{MRI/CT registration. The zoomed-in region in Figure~\ref{fig:ct_mr_roi_zoomed_in_view} shows good alignment of bone tip between MRI and registered CT images using proposed method. The False-color overlays shown in Figure~\ref{fig:mr-ct-overlay} demonstrates spatial correspondence between fixed (MR in red) and registered (CT in green) volumes for a representative slice. Post registration there is good alignment of body contour. For both examples, proposed methods seems to perform slightly better than ANTs.}
	\label{fig:ct_mr_results}
	
	\begin{subfigure}{0.75\textwidth}
		\centering
		\includegraphics[width=\linewidth]{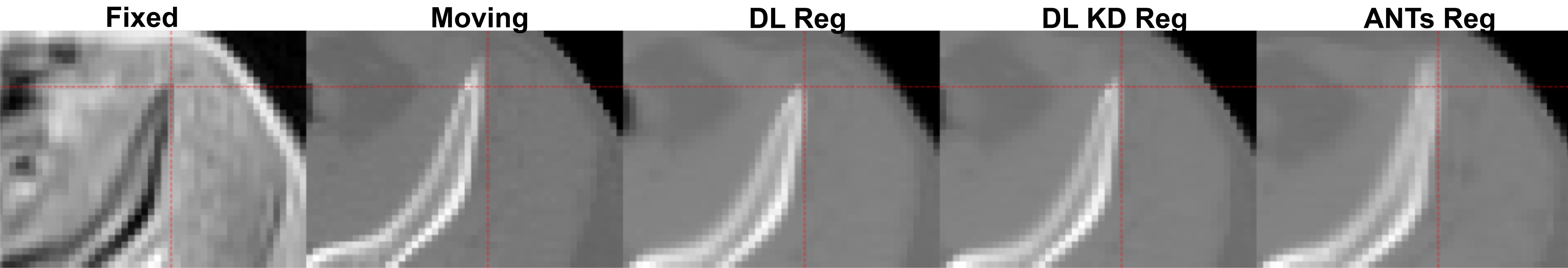}
		\caption{}%
		\label{fig:ct_mr_roi_zoomed_in_view}
	\end{subfigure}
	
	\medskip
	
	\begin{subfigure}{0.85\textwidth}
		\centering
		\includegraphics[width=\linewidth]{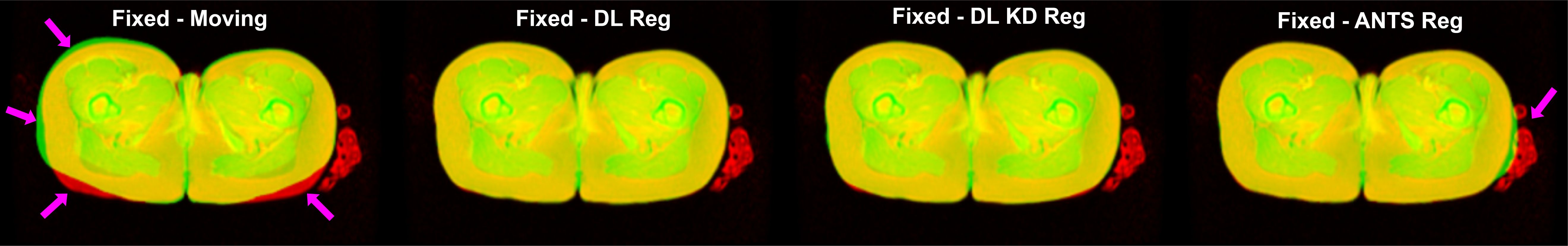}
		\caption{}%
		\label{fig:mr-ct-overlay}
	\end{subfigure}
\end{figure}

Figure~\ref{fig:ct_mr_results} illustrates qualitative results of registration using the proposed TTO framework. The comparison in Figure~\ref{fig:ct_mr_roi_zoomed_in_view} highlights accurate alignment of anatomical structures (red crosshair at bone tip) across modalities while preserving modality-specific characteristics. 
The MRI-CT image overlay for another slice is shown in Figure~\ref{fig:mr-ct-overlay}. The fixed image is shown in red and the registered image in green (blue channel suppressed). Regions of accurate correspondence appear yellow (red + green), while red-only and green-only regions indicate anatomical structures present exclusively in the fixed or registered image, respectively. It demonstrates fine alignment post proposed registration method (both body contour and internal tissues).  
PMMs for few slices from this data are shown in Appendix~\ref{sec:appendix-MT Maps CT-MR} (Fig.~\ref{fig:ct-mr-mi-maps}). These maps were computed using 3D patches of size $16 \times 16 \times 16$ voxels with a stride of 4. Mean PMM values across all slices are: Fixed–Moving = 0.0986, Fixed–DL Reg = 0.1797, Fixed–KD Reg = 0.1680, and Fixed–ANTs Reg = 0.1742.

\subsection{Pre- and Post-Contrast CT Registration}
The results of the experiments discussed in Section~\ref{subsubsec:prepostct} are discussed here. The corresponding organ segmentation overlaps (performed using~\cite{wasserthal2023totalsegmentator}) are considered for comparison: lung segmentation for \textit{Data-1} and liver segmentation for \textit{Data-2}, \textit{Data-3}, and \textit{Data-4}. Dice and IoU scores pre and post registration using proposed DL registration method (with and without KD), and ANTs are shown in Table~\ref{tab:dice_iou_scores}. Post registration, the dice and IoU scores have improved across cases. The performance of the proposed method is similar to that of ANTs. Refer to Appendix~\ref{sec:appendix-pre-post-contrast-CT} Figure~\ref{fig:pre-post-contrast-qual-results} for additional qualitative assessments.

\begin{table}[htb]
	\centering
	\resizebox{0.95\textwidth}{!}{%
		\renewcommand{\arraystretch}{1.2}
		\begin{tabular}{ccccccccc}
			\toprule
			\multirow{2}{*}{\textbf{Data}} & \multicolumn{4}{c}{\textbf{Dice Score}} & \multicolumn{4}{c}{\textbf{IoU Score}} \\
			\cmidrule(lr){2-5} \cmidrule(lr){6-9}
			& F-M & F-R & F-R-KD & F-AR & F-M & F-R & F-R-KD & F-AR \\
			\midrule
			Data-1 & 0.9834 & 0.9838 & \cellcolor{gray!30}0.9928 & 0.9844 & 0.9673 & 0.9681 & \cellcolor{gray!30}0.9857 & 0.9694 \\
			Data-2 & 0.9393 & 0.9737 & \cellcolor{gray!30}0.9744 & 0.9733 & 0.8856 & 0.9488 & \cellcolor{gray!30}0.9500 & 0.9480 \\
			Data-3 & 0.9576 & \cellcolor{gray!30}0.9755 & 0.9691 & 0.9526 & 0.9186 & \cellcolor{gray!30}0.9523 & 0.9401 & 0.9095 \\
			Data-4 & 0.8096 & 0.9675 & 0.9519 & \cellcolor{gray!30}0.9724 & 0.6801 & 0.9371 & 0.9082 & \cellcolor{gray!30}0.9463 \\
			\bottomrule
	\end{tabular}}
	\caption{Comparison of Dice and IoU scores for primary organ segmentation from pre- and post-contrast CT datasets are shown here. Evaluations are performed across four settings: Fixed-Moving (F-M), Fixed-DL Reg (F-R), Fixed-DL Reg KD (F-R-KD), and Fixed-ANTs registered (F-AR).}
	\label{tab:dice_iou_scores}
\end{table}
\subsection{Brain MRI Registration}
\begin{figure}[htb]
	\centering
	\includegraphics[width=1.0\textwidth]{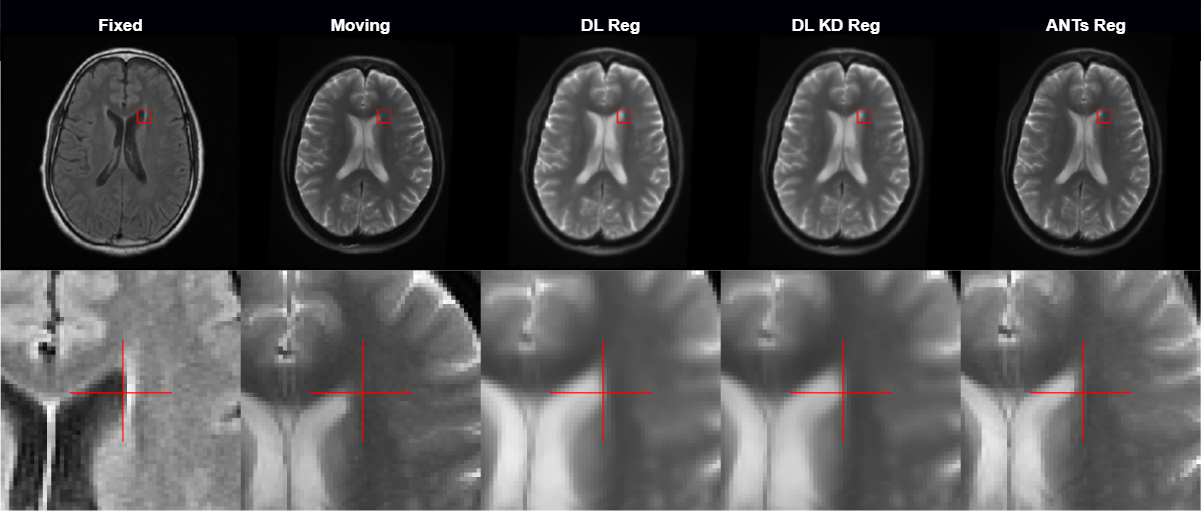}
	\caption{Qualitative comparison of inter-subject cross-contrast MRI registration.The zoomed-in region within the red bounding box highlights accurate alignment of the frontal horn of the lateral ventricles post registration.}
	\label{fig:Cross-Contrast Brain MR zoomed ROI}
\end{figure}

The results of the experiments discussed in Section~\ref{subsubsec:expneuroreg} are discussed here.
Results of an inter-subject cross-contrast registration is shown in Figure~\ref{fig:Cross-Contrast Brain MR zoomed ROI}. The red square zooms into region where ventricle ends are present in the 2D slice, and have sufficient misalignment post affine registration. The structural alignment is restored post registration using the proposed method. Based on the qualitative analysis for this example, proposed method performs slightly better than ANTs registration.
For quantitative evaluation for all cases, PMM was computed for each fixed–moving/registered pair (see Appendix~\ref{sec:appendix-patchwiseMI} for visual results). Mean PMM values were computed for each case inside the brain mask. Table~\ref{tab:mi_stats_brain_mr} summarizes results for three scenarios: \textit{Same-subject, Cross-contrast} ($n=8$), \textit{Inter-subject, Same-contrast} ($n=24$), and \textit{Inter-subject, Cross-contrast} ($n=24$). Comparisons include the unregistered moving image (F--M), DL-registered outputs using the dense model (F--R) and KD-based model (F--R--KD), and ANTs registration (F--AR).
Across all evaluation combinations, proposed registration method improved PMM values relative to the unregistered baseline. The DL Reg model achieved the highest PMM. The DL Reg KD model closely follows its performance. \textit{Same-subject, Cross-contrast} yielded the highest post-registration PMM with low variability, while \textit{Inter-subject, Cross-contrast} showed a relatively lower PMM, reflecting the increased complexity of inter-subject and cross-contrast alignment. Otherwise, one must note that same-contrast image pairs will have higher PMM values by virtue of contrast matching.
Additional qualitative comparisons are shown in Appendix~\ref{sec:appendix-MT Maps Brain MR}.


\begin{table}[htb]
	\centering
	\resizebox{0.95\textwidth}{!}{%
		\begin{tabular}{lcccccccc}
			\toprule
			& \multicolumn{2}{c}{F-M} & \multicolumn{2}{c}{F-R} & \multicolumn{2}{c}{F-R-KD} & \multicolumn{2}{c}{F-AR} \\
			\cmidrule(r){2-3}\cmidrule(r){4-5}\cmidrule(r){6-7}\cmidrule(r){8-9}
			Category & Mean & Std & Mean & Std & Mean & Std & Mean & Std \\
			\midrule
			Same-subject, Cross-contrast
			& 0.1677 & 0.0108 & \cellcolor{gray!30}0.2078 & 0.0135 & 0.2032 & \cellcolor{gray!30}0.0138 & 0.1861 & 0.0209 \\
			Inter-subject, Same-contrast
			& 0.0887 & 0.0114 & \cellcolor{gray!30}0.1707 & \cellcolor{gray!30}0.0192 & 0.1618 & 0.0190 & 0.1389 & 0.0151 \\
			Inter-subject, Cross-contrast
			& 0.0827 & 0.0077 & \cellcolor{gray!30}0.1372 & \cellcolor{gray!30}0.0096 & 0.1291 & 0.0092 & 0.1120 & 0.0092 \\
			\bottomrule
		\end{tabular}
	}
	\caption{Mean and standard deviation of PMMs across categories (values computed inside brain mask).}
	\label{tab:mi_stats_brain_mr}
\end{table}
\section{Conclusion}
In this work we proposed a DL-based registration framework which generalizes well across applications. Even though test time optimization concepts are used, by using model miniaturization approaches, the proposed method on an average takes half the time (or lesser) taken by popular analytical registration libraries. The ability of the proposed method to align structures across modalities and anatomies have been demonstrated using qualitative and quantitative measures.
\section*{Acknowledgments}
The authors acknowledge the use of data from the following sources:
	
	\begin{itemize}
		\item The Cancer Genome Atlas Uterine Corpus Endometrial Carcinoma Collection (TCGA-UCEC) \cite{Erickson2016TCGAUCEC}. The results published here are in whole or part based upon data generated by the TCGA Research Network: \url{http://cancergenome.nih.gov/}.
		
		\item VA Research Precision Oncology Program – APOLLO (VAREPOP-APOLLO) \cite{APOLLO2024VAREPOP}. Data used in this publication were generated by the Veterans Health Administration’s Research for Precision Oncology Program and the Applied Proteogenomics Organizational Learning and Outcomes (APOLLO) Research Network.
\end{itemize}

\bibliographystyle{unsrtnat}
\bibliography{references}  

\begin{thebibliography}{20}
\providecommand{\natexlab}[1]{#1}
\providecommand{\url}[1]{\texttt{#1}}
\expandafter\ifx\csname urlstyle\endcsname\relax
  \providecommand{\doi}[1]{doi: #1}\else
  \providecommand{\doi}{doi: \begingroup \urlstyle{rm}\Url}\fi

\bibitem[Hill et~al.(2001)Hill, Batchelor, Holden, and Hawkes]{hill2001medical}
Derek~LG Hill, Philipp~G Batchelor, Mark Holden, and David~J Hawkes.
\newblock Medical image registration.
\newblock \emph{Physics in Medicine \& Biology}, 46\penalty0 (3):\penalty0 R1,
  2001.

\bibitem[Maintz and Viergever(1998)]{maintz1998survey}
JB~Antoine Maintz and Max~A Viergever.
\newblock A survey of medical image registration.
\newblock \emph{Medical Image Analysis}, 2\penalty0 (1):\penalty0 1--36, 1998.

\bibitem[Sotiras et~al.(2013)Sotiras, Davatzikos, and
  Paragios]{sotiras2013deformable}
Aristeidis Sotiras, Christos Davatzikos, and Nikos Paragios.
\newblock Deformable medical image registration: a survey.
\newblock \emph{IEEE Transactions on Medical Imaging}, 32\penalty0
  (7):\penalty0 1153--1190, 2013.

\bibitem[Oliveira and Tavares(2014)]{oliveira2014medical}
Francisco~PM Oliveira and Jo{\~a}o Manuel~RS Tavares.
\newblock Medical image registration: a review.
\newblock \emph{Computer Methods in Biomechanics and Biomedical Engineering},
  17\penalty0 (2):\penalty0 73--93, 2014.

\bibitem[Avants et~al.(2009)Avants, Tustison, and Song]{avants2009ants}
Brian~B. Avants, Nicholas~J. Tustison, and Gang Song.
\newblock Advanced normalization tools (ants).
\newblock \emph{Insight Journal}, pages 1--35, 2009.
\newblock URL \url{https://github.com/ANTsX/ANTs}.
\newblock https://www.insight-journal.org/browse/publication/681.

\bibitem[Balakrishnan et~al.(2019)Balakrishnan, Zhao, Sabuncu, Guttag, and
  Dalca]{balakrishnan2019tmi}
Guha Balakrishnan, Amy Zhao, Mert Sabuncu, John Guttag, and Adrian~V. Dalca.
\newblock Voxelmorph: A learning framework for deformable medical image
  registration.
\newblock \emph{IEEE TMI: Transactions on Medical Imaging}, 38:\penalty0
  1788--1800, 2019.

\bibitem[Reithmeir et~al.(2026)Reithmeir, Spieker, Sideri-Lampretsa, Rueckert,
  Schnabel, and Zimmer]{REITHMEIR2026103854}
Anna Reithmeir, Veronika Spieker, Vasiliki Sideri-Lampretsa, Daniel Rueckert,
  Julia~A. Schnabel, and Veronika~A. Zimmer.
\newblock From model based to learned regularization in medical image
  registration: A comprehensive review.
\newblock \emph{Medical Image Analysis}, 108:\penalty0 103854, 2026.
\newblock ISSN 1361-8415.
\newblock \doi{https://doi.org/10.1016/j.media.2025.103854}.
\newblock URL
  \url{https://www.sciencedirect.com/science/article/pii/S1361841525004001}.

\bibitem[Fu et~al.(2020)Fu, Brown, Saeed, Casamitjana, Baum, Delaunay, Yang,
  Grimwood, Min, Blumberg, Iglesias, Barratt, Bonmati, Alexander, Clarkson,
  Vercauteren, and Hu]{Fu2020}
Yunguan Fu, Nina~Montaña Brown, Shaheer~U. Saeed, Adrià Casamitjana, Zachary
  M.~C. Baum, Rémi Delaunay, Qianye Yang, Alexander Grimwood, Zhe Min,
  Stefano~B. Blumberg, Juan~Eugenio Iglesias, Dean~C. Barratt, Ester Bonmati,
  Daniel~C. Alexander, Matthew~J. Clarkson, Tom Vercauteren, and Yipeng Hu.
\newblock Deepreg: a deep learning toolkit for medical image registration.
\newblock \emph{Journal of Open Source Software}, 5\penalty0 (55):\penalty0
  2705, 2020.
\newblock \doi{10.21105/joss.02705}.
\newblock URL \url{https://doi.org/10.21105/joss.02705}.

\bibitem[Zhang et~al.(2021)Zhang, Wu, Gach, Li, and Yang]{zhang2021groupregnet}
Yunlu Zhang, Xue Wu, H~Michael Gach, Harold Li, and Deshan Yang.
\newblock Groupregnet: A groupwise one-shot deep learning-based 4d image
  registration method.
\newblock \emph{Physics in Medicine \& Biology}, 66\penalty0 (4):\penalty0
  045030, 2021.

\bibitem[Ma et~al.(2024)Ma, Zhang, Li, and Wen]{ma2024iirpnet}
Tai Ma, Suwei Zhang, Jiafeng Li, and Ying Wen.
\newblock Iirp-net: Iterative inference residual pyramid network for enhanced
  image registration.
\newblock In \emph{Proceedings of the IEEE/CVF Conference on Computer Vision
  and Pattern Recognition (CVPR)}, pages 11546--11555, 2024.

\bibitem[Demir et~al.(2024)Demir, Tian, Greer, Kwitt, Vialard, Estepar, Bouix,
  Rushmore, Ebrahim, and Niethammer]{demir2024multigradicon}
Basar Demir, Lin Tian, Thomas~Hastings Greer, Roland Kwitt, Francois-Xavier
  Vialard, Raul San~Jose Estepar, Sylvain Bouix, Richard~Jarrett Rushmore,
  Ebrahim Ebrahim, and Marc Niethammer.
\newblock multigradicon: A foundation model for multimodal medical image
  registration.
\newblock \emph{arXiv preprint arXiv:2408.00221}, 2024.

\bibitem[Tian et~al.(2024)Tian, Greer, Kwitt, Vialard, Estepar, Bouix,
  Rushmore, and Niethammer]{tian2024unigradicon}
Lin Tian, Hastings Greer, Roland Kwitt, Francois-Xavier Vialard, Raul San~Jose
  Estepar, Sylvain Bouix, Richard Rushmore, and Marc Niethammer.
\newblock unigradicon: A foundation model for medical image registration.
\newblock \emph{arXiv preprint arXiv:2403.05780}, 2024.

\bibitem[Hoffmann et~al.(2022)Hoffmann, Billot, Greve, Iglesias, Fischl, and
  Dalca]{Hoffmann_2022}
Malte Hoffmann, Benjamin Billot, Douglas~N. Greve, Juan~Eugenio Iglesias, Bruce
  Fischl, and Adrian~V. Dalca.
\newblock Synthmorph: Learning contrast-invariant registration without acquired
  images.
\newblock \emph{IEEE Transactions on Medical Imaging}, 41\penalty0
  (3):\penalty0 543–558, March 2022.
\newblock ISSN 1558-254X.
\newblock \doi{10.1109/tmi.2021.3116879}.
\newblock URL \url{http://dx.doi.org/10.1109/TMI.2021.3116879}.

\bibitem[Hinton et~al.(2015)Hinton, Vinyals, and Dean]{hinton2015distilling}
Geoffrey Hinton, Oriol Vinyals, and Jeff Dean.
\newblock Distilling the knowledge in a neural network.
\newblock \emph{arXiv preprint arXiv:1503.02531}, 2015.
\newblock URL \url{https://arxiv.org/abs/1503.02531}.

\bibitem[{\c{C}}i{\c{c}}ek et~al.(2016){\c{C}}i{\c{c}}ek, Abdulkadir, Lienkamp,
  Brox, and Ronneberger]{cciccek20163d}
{\"O}zg{\"u}n {\c{C}}i{\c{c}}ek, Ahmed Abdulkadir, Soeren~S Lienkamp, Thomas
  Brox, and Olaf Ronneberger.
\newblock 3d u-net: learning dense volumetric segmentation from sparse
  annotation.
\newblock In \emph{International conference on medical image computing and
  computer-assisted intervention}, pages 424--432. Springer, 2016.

\bibitem[Bakr et~al.(2017)Bakr, Gevaert, Echegaray, Ayers, Zhou, Shafiq, Zheng,
  Zhang, Leung, Kadoch, Shrager, Quon, Rubin, Plevritis, and
  Napel]{Bakr2017NSCLC}
S.~Bakr, O.~Gevaert, S.~Echegaray, K.~Ayers, M.~Zhou, M.~Shafiq, H.~Zheng,
  W.~Zhang, A.~Leung, M.~Kadoch, J.~Shrager, A.~Quon, D.~Rubin, S.~Plevritis,
  and S.~Napel.
\newblock Data for nsclc radiogenomics (version 4) [data set], 2017.
\newblock URL \url{https://doi.org/10.7937/K9/TCIA.2017.7hs46erv}.
\newblock Acknowledgment: Publications using data from this program are
  requested to include the following statement: "The results <published or
  shown> here are in whole or part based upon data generated by the TCGA
  Research Network: http://cancergenome.nih.gov/.".

\bibitem[Erickson et~al.(2016)Erickson, Mutch, Lippmann, and
  Jarosz]{Erickson2016TCGAUCEC}
B.~J. Erickson, D.~Mutch, L.~Lippmann, and R.~Jarosz.
\newblock The cancer genome atlas uterine corpus endometrial carcinoma
  collection (tcga-ucec) (version 4) [data set], 2016.
\newblock URL \url{https://doi.org/10.7937/K9/TCIA.2016.GKJ0ZWAC}.
\newblock Acknowledgment: Publications using data from this program are
  requested to include the following statement: "The results <published or
  shown> here are in whole or part based upon data generated by the TCGA
  Research Network: http://cancergenome.nih.gov/.".

\bibitem[for Precision Oncology~Program et~al.(2024)for Precision
  Oncology~Program, the Applied Proteogenomics Organizational~Learning, and
  Network]{APOLLO2024VAREPOP}
The~Research for Precision Oncology~Program, the Applied Proteogenomics
  Organizational~Learning, and Outcomes (APOLLO)~Research Network.
\newblock Va research precision oncology program – apollo (varepop-apollo)
  (version 3) [data set], 2024.
\newblock URL \url{https://doi.org/10.7937/ghkn-md15}.
\newblock Acknowledgment: Data used in this publication were generated by the
  Veterans Health Administration’s Research for Precision Oncology Program
  and the Applied Proteogenomics Organizational Learning and Outcomes (APOLLO)
  Research Network.

\bibitem[Moawad et~al.(2021)Moawad, Fuentes, Morshid, Khalaf, Elmohr, Abusaif,
  Hazle, Kaseb, Hassan, Mahvash, Szklaruk, Qayyom, and Elsayes]{Moawad2021HCC}
A.~W. Moawad, D.~Fuentes, A.~Morshid, A.~M. Khalaf, M.~M. Elmohr, A.~Abusaif,
  J.~D. Hazle, A.~O. Kaseb, M.~Hassan, A.~Mahvash, J.~Szklaruk, A.~Qayyom, and
  K.~Elsayes.
\newblock Multimodality annotated hcc cases with and without advanced imaging
  segmentation [data set], 2021.
\newblock URL \url{https://doi.org/10.7937/TCIA.5FNA-0924}.

\bibitem[Wasserthal et~al.(2023)Wasserthal, Breit, Meyer, and
  et~al.]{wasserthal2023totalsegmentator}
Jakob Wasserthal, Holger Breit, Martin Meyer, and et~al.
\newblock Totalsegmentator: Robust segmentation of 104 anatomical structures in
  ct images.
\newblock \emph{Radiology: Artificial Intelligence}, 5\penalty0 (5):\penalty0
  e230024, 2023.
\newblock \doi{10.1148/ryai.230024}.
\newblock URL \url{https://github.com/wasserth/TotalSegmentator}.

\end{thebibliography}
\appendix
\section{Multi-Axis Mutual Information Similarity Loss}
\label{sec:appendix-MultiAxisMI}
We adopt the Multi-Axis Mutual Information (MultiAxisMI) approach instead of a full 3D Mutual Information (MI) loss. The similarity loss is defined as:
\begin{equation}
	\mathcal{L}_{\text{MultiAxisMI}} = \frac{1}{D+H+W} \Bigg( \sum_{d=1}^{D} \mathcal{M}(I_f^{(d)}, I_m^{(d)} \circ \phi) + \sum_{h=1}^{H} \mathcal{M}(I_f^{(h)}, I_m^{(h)} \circ \phi) + \sum_{w=1}^{W} \mathcal{M}(I_f^{(w)}, I_m^{(w)} \circ \phi) \Bigg),
	\label{eq:multi_axis_mi}
\end{equation}
where:
\begin{itemize}
	\item $\mathcal{M}(\cdot,\cdot)$ denotes the mutual information between corresponding slices.
	\item $I_f^{(d)}, I_m^{(d)}$ represent slices along the depth axis, and similarly for height ($h$) and width ($w$).
	\item $D, H, W$ are the number of slices along each axis.
\end{itemize}
This formulation ensures robustness to local intensity variations and captures anatomical alignment across all three dimensions.

\section{Normalized Cross-Correlation (NCC) Loss}
\label{sec:appendix-ncc}

The Normalized Cross-Correlation (NCC) loss used in our experiments is defined as:
\begin{equation}
	\mathcal{L}_{\text{NCC}}(I_f, I_m \circ \phi) =
	- \frac{\sum_{x} (I_f(x) - \bar{I}_f)(I_m(\phi(x)) - \bar{I}_m)}
	{\sqrt{\sum_{x} (I_f(x) - \bar{I}_f)^2} \, \sqrt{\sum_{x} (I_m(\phi(x)) - \bar{I}_m)^2}},
	\label{eq:ncc_loss}
\end{equation}
where $\bar{I}_f$ and $\bar{I}_m$ denote the mean intensities of the fixed and warped moving images, respectively.

\section{Patchwise Mutual Information over ROI or Full Volume}
\label{sec:appendix-patchwiseMI}

To evaluate local registration quality, we compute a \emph{patchwise mutual information (MI) map} over a specified spatial domain, which can be either a predefined anatomical region of interest (ROI) or the entire image volume.

Let $F$ and $M$ denote the fixed and moving (or registered) 3D images with voxel intensities
$F(\mathbf{r}), M(\mathbf{r})$ at spatial location $\mathbf{r} \in \Omega \subset \mathbb{Z}^3$.
Define a binary mask $\mathcal{D} \subseteq \Omega$ representing the domain of interest (e.g., ROI or full volume). A voxel belongs to the domain iff $\mathbf{r} \in \mathcal{D}$.

We define a family of overlapping cubic patches $\{P_k\}_{k=1}^K$ covering $\Omega$. For each patch $P_k$, we consider its intersection with the domain:
\[
P_k^\mathcal{D} = P_k \cap \mathcal{D}.
\]
If $|P_k^\mathcal{D}|$ exceeds a minimal sample threshold, we estimate the empirical joint intensity distribution $p_{F,M}^{(k)}(f,m)$ over voxels in $P_k^\mathcal{D}$, and compute the local mutual information:
\[
I_k = I(F;M \mid P_k^\mathcal{D}) = \sum_{f,m} p_{F,M}^{(k)}(f,m)\,
\log \frac{p_{F,M}^{(k)}(f,m)}{p_{F}^{(k)}(f)\,p_{M}^{(k)}(m)}.
\]

Assigning $I_k$ to all voxels in $P_k^\mathcal{D}$ and averaging over overlapping patches produces a continuous MI field:
\[
\mathrm{MI}_{\mathcal{D}}(\mathbf{r}) = \frac{1}{N(\mathbf{r})} \sum_{k:\,\mathbf{r} \in P_k^\mathcal{D}} I_k,\quad \mathbf{r} \in \mathcal{D},
\]
where $N(\mathbf{r})$ is the number of domain-containing patches covering $\mathbf{r}$.

This spatial MI map localizes registration quality: higher values indicate stronger statistical dependence and better alignment. Patchwise aggregation reveals heterogeneous performance (e.g., cortical vs. deep structures), which a single global MI would conceal. Mean statistics of $\mathrm{MI}_{\mathcal{D}}$ before and after registration (or across methods) provide interpretable improvement scores (e.g., $\Delta \mathrm{MI}_{\mathcal{D}}$).

\section{Qualitative analysis of DCE MRI registration}\label{sec:appendixQualityDCE}
\begin{figure}[htb]
	\centering
	\includegraphics[width=0.7\textwidth]{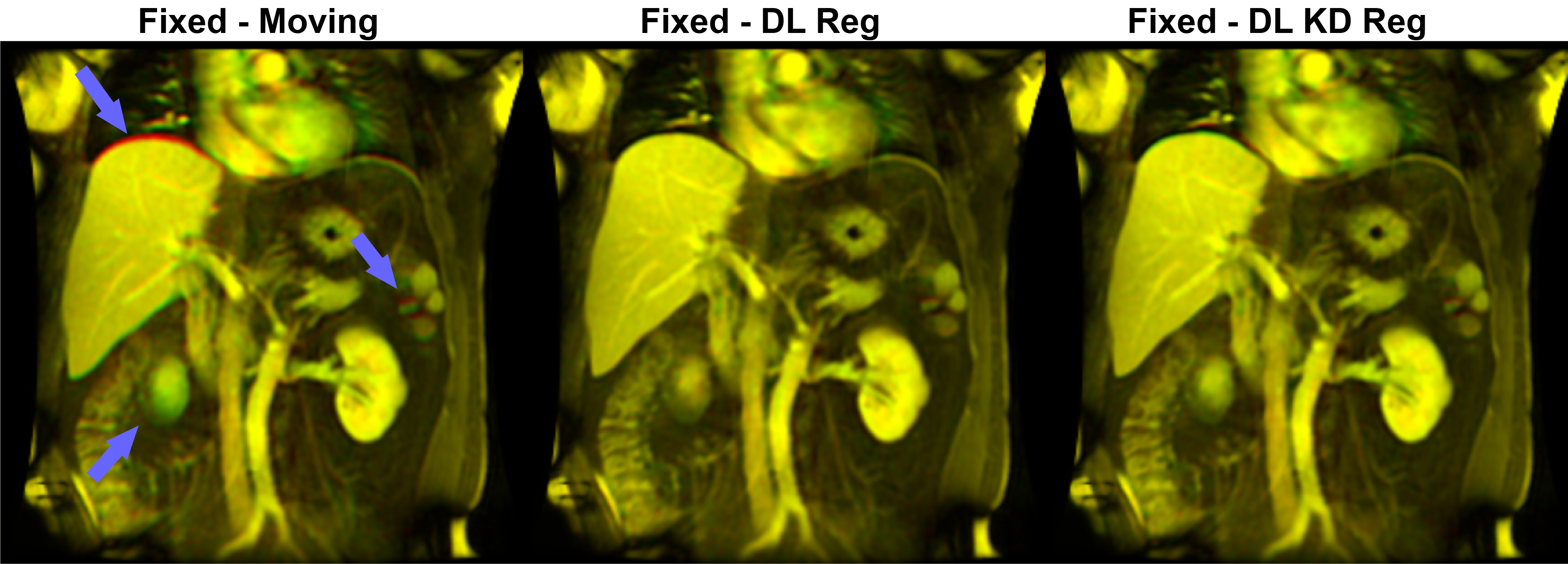}
	\caption{Qualitative overlays for liver DCE-MRI registration.}
	\label{fig:overlayDCEMRI}
\end{figure}

\section{Qualitative analysis of MRI-CT Multi-Modal Registration}
\begin{figure}[htb]
	\centering
	\includegraphics[width=0.85\textwidth]{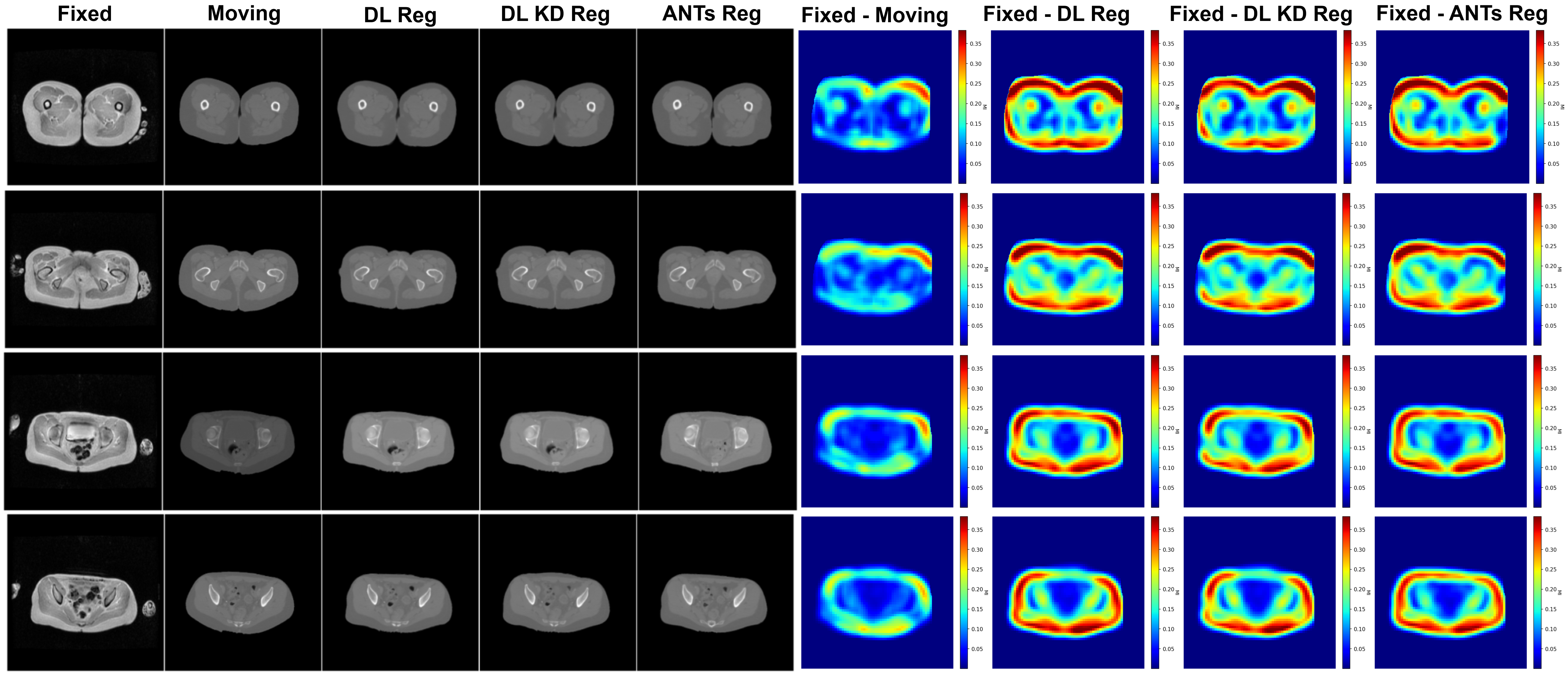}
	\caption{MRI/CT Registration -  patchwise MI Maps of several slices}
	\label{fig:ct-mr-mi-maps}
\end{figure}

Figure ~\ref{fig:ct-mr-mi-maps} illustrates patch-wise MI maps for several slices, computed using a patch size of $16 \times 16 \times 16$ voxels and a stride of 4, following the approach described in Section~\ref{sec:appendix-patchwiseMI}.
\label{sec:appendix-MT Maps CT-MR}


\section{Qualitative analysis of Cross-Contrast Brain-MR Registration}
\begin{figure}[htb]
	\centering
	\includegraphics[width=0.85\textwidth]{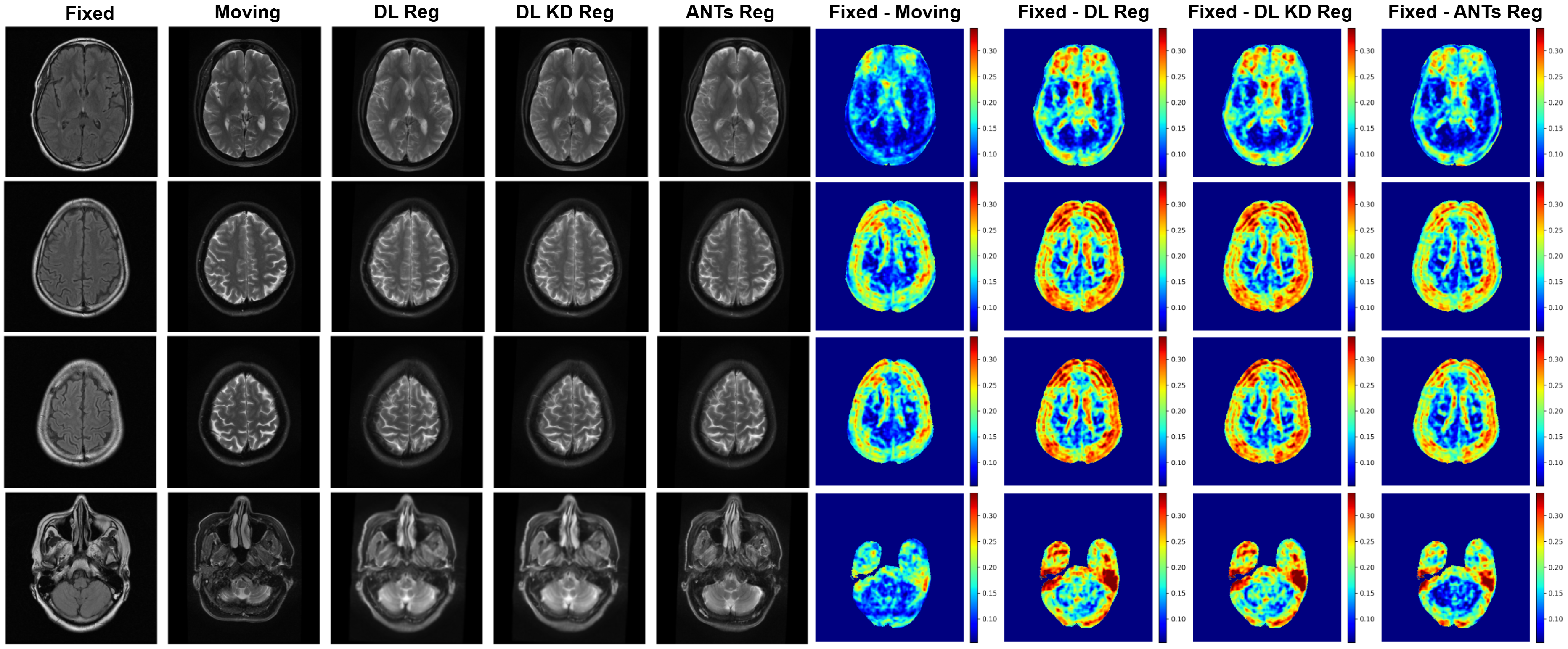}
	\caption{Inter Subject Cross Contrast Brain MRI Registration -  patchwise MI Maps of several slices}
	\label{fig:MI Maps of Cross-Contrast Brain MR}
\end{figure}

Figure~\ref{fig:MI Maps of Cross-Contrast Brain MR} illustrates patch-wise MI maps for several slices, computed using a patch size of $16 \times 16 \times 16$ voxels and a stride of 4, (refer Appendix~\ref{sec:appendix-patchwiseMI}. The computation is performed within the brain mask to ensure anatomical relevance.
\label{sec:appendix-MT Maps Brain MR}

\begin{figure}[htb]
	\centering
	\includegraphics[width=0.8\textwidth]{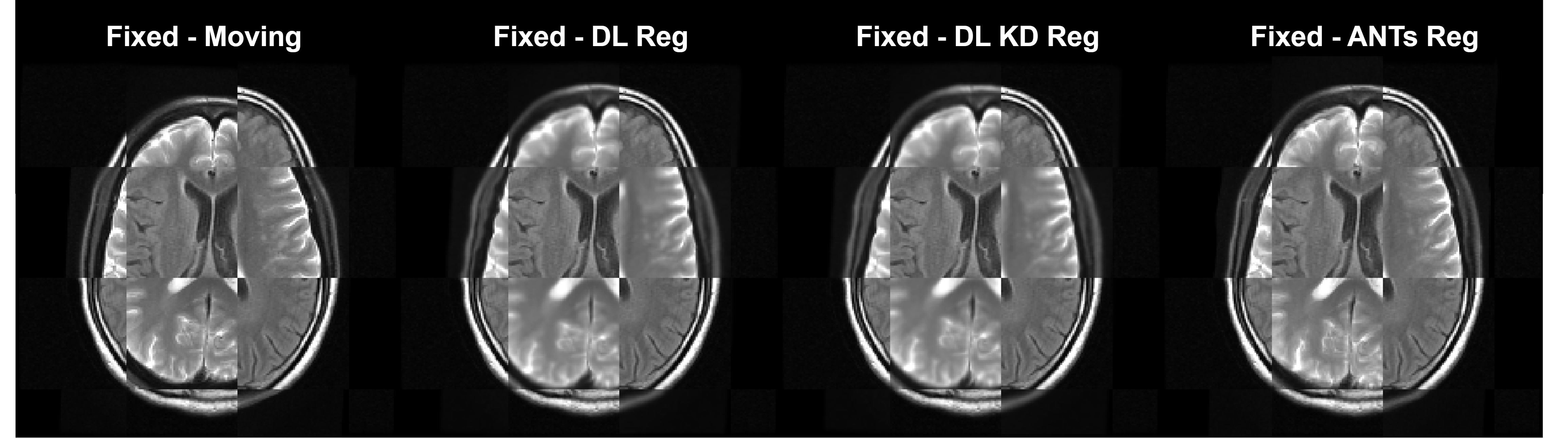}
	\caption{Checkerboard overlays illustrating local alignment between fixed and registered volumes for a slice. Alternating tiles display either the fixed image or the corresponding registered image, facilitating localized visual assessment of registration accuracy. Smooth transitions across tile boundaries indicate accurate local registration, whereas visible seams or discontinuities reveal residual misalignment.}
	\label{fig:Cross-Contrast Brain MR checkerboard overlay}
\end{figure}

Figure~\ref{fig:Cross-Contrast Brain MR checkerboard overlay} shows checkerboard overlays of inter-subject cross-contrast brain MR images before and after registration, visually emphasizing the improved alignment achieved through the proposed method.

\section{Pre and Post Contrast CT Qualitative evaluation}
\begin{figure}[htb]
	\centering
	\caption{Pre- and post-contrast CT registration. Figure~\ref{fig:pre-post-contrast-image-slice} illustrates accurate alignment of the liver dome between the pre-contrast CT and the registered post-contrast CT using the proposed method. The segmentation overlays in Figure~\ref{fig:pre-post-contrast-mask-overlays} further confirm spatial correspondence between the fixed and registered volumes for a representative slice, demonstrating improved alignment of liver segmentation after registration.}
	\label{fig:pre-post-contrast-qual-results}
	
	\begin{subfigure}{0.8\textwidth}
		\centering
		\includegraphics[width=\linewidth]{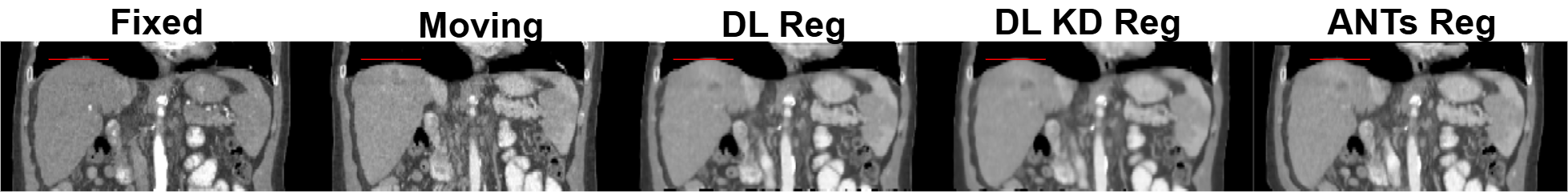}
		\caption{}%
		\label{fig:pre-post-contrast-image-slice}
	\end{subfigure}
	
	\medskip
	
	\begin{subfigure}{0.8\textwidth}
		\centering
		\includegraphics[width=\linewidth]{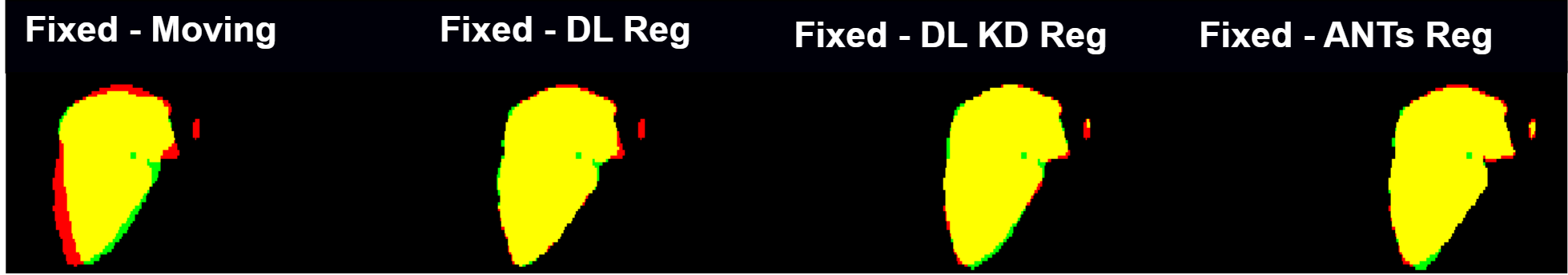}
		\caption{}%
		\label{fig:pre-post-contrast-mask-overlays}
	\end{subfigure}
\end{figure}

For Data-4 (Section~\ref{subsubsec:prepostct}), Figure~\ref{fig:pre-post-contrast-qual-results} presents qualitative results of pre- and post-contrast CT registration, illustrating improved liver alignment and segmentation consistency after registration.

\label{sec:appendix-pre-post-contrast-CT}
\clearpage
\section{Knowledge Distillation Framework}
\label{sec:appendix-KD-framework}
\begin{figure}[htb]
	\centering
	\includegraphics[width=0.75\textwidth]{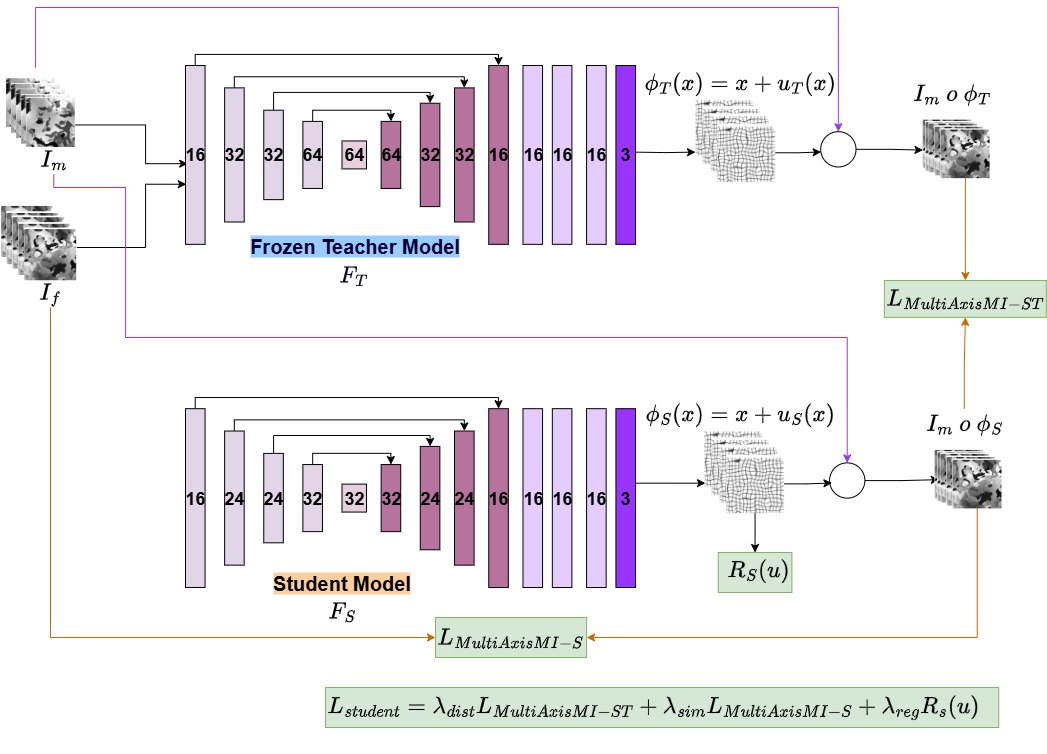}
	\caption{Knowledge Distillation framework - The trained and frozen teacher model guides the student using output based distillation, image similarity and smoothness losses.}
	\label{fig:KD framework}
\end{figure}






\end{document}